\documentclass[aps,10pt,pra,twocolumn,floatfix,floats,showpacs,superscriptaddress,raggedbottom]{revtex4-1}
\usepackage{graphicx,latexsym}
\usepackage{dcolumn}
\usepackage{amsmath}
\usepackage{amssymb,bm}
\usepackage{braket}
\usepackage{natbib}

\usepackage[normalem]{ulem}

\def\sec#1{Sec.\ \ref{#1}}
\def\eq#1{Eq.\ (\ref{#1})}
\def\fig#1{Fig.\ \ref{#1}}

\begin{document}

\title{The photocurrent generated by photon replica states \\ 
     of an off-resonantly coupled dot-cavity system}

\author{Nzar Rauf Abdullah}
\email{nzar.r.abdullah@gmail.com}
\affiliation{Physics Department, College of Science, 
             University of Sulaimani, Kurdistan Region, Iraq}
\affiliation{Komar Research Center, Komar University of Science and Technology, Sulaimani, Iraq}

\author{Chi-Shung Tang}
\affiliation{Department of Mechanical Engineering,
  National United University, 1, Lienda, Miaoli 36003, Taiwan}

\author{Andrei Manolescu}
\affiliation{Reykjavik University, School of Science and Engineering,
              Menntavegur 1, IS-101 Reykjavik, Iceland}

\author{Vidar Gudmundsson}
\email{vidar@hi.is}
 \affiliation{Science Institute, University of Iceland,
        Dunhaga 3, IS-107 Reykjavik, Iceland}

%

\begin{abstract}
Transport properties of a quantum dot coupled to a photon cavity are investigated using a quantum master equation 
in the steady-state regime.  In the off-resonance regime, when the photon energy is smaller than the energy spacing 
between the lowest electron states of the quantum dot, we calculate the current that is generated by photon replica 
states as the electronic system is pumped with photons.
Tuning the electron-photon coupling strength, the photocurrent can be enhanced by the influences of the photon polarization, 
and the cavity-photon coupling strength of the environment. We show that the current generated through the photon 
replicas is very sensitive on the photon polarization, but it is not strongly dependent on the average number of photons 
in the environment.
\end{abstract}



\maketitle

%
%

\section{Introduction}\label{Sec:Introduction}

The light-matter interaction in nanoscale systems is one of the most fundamental 
and interesting topic of modern nanodevices~\cite{Delbecq2013,PhysRevB.78.125308},
especially if the light consists of few photons. In this case the light must be treated as fully 
quantized~\cite{Giannelli_2018, Kreinberg2018,Giannelli_2018}. 
Few photons interacting with a quantized electronic system, both in weak and strong coupling 
regimes, are very attractive for fundamental research and applications of nanotechnology~\cite{PhysRevB.87.115419,PhysRevB.91.205417}.
For instance, an unconventional photon blockade are observed in a quantum dot (QD) weakly coupled to a QED cavity, in which 
a photon blockade effect can control the efficiency of single photon sources~\cite{PhysRevLett.121.043601}.
In addition, in the strong coupling limit a quantum cavity coupled to a double quantum dot (DQD) system
offers the capability of a coherent spectroscopy of a DQD qubit in the dispersive regime~\cite{PhysRevX.7.011030}.

The coupling strength between an electronic system and a photon field in a cavity, $g_\gamma¸$, can be compared to the 
coupling strength of the cavity to the environment, $\kappa$~\cite{doi:10.1063/1.3294298}. 
The system is said to be in the strong coupling regime if $g_{\gamma} > \kappa$.
Note though that in addition to this condition the strength of the interaction could also be 
compared to a characteristic energy spacing of the electron system \cite{FriskKockum2019}. 
In the strong coupling regime, several interesting phenomena have been observed such as 
photon-induced tunneling in a vacuum Rabi split two level quantum dot system~\cite{Faraon2008} and 
photon blockade in the presence of an effective photon-photon interactions in a qubit system~\cite{PhysRevLett.106.243601}.
Recently, it has been proposed that the entanglement properties of the cavity-photons with electrons in a strong coupling regime 
can be used to mediate non-Coulombic entanglement between two distant electrons~\cite{Ofer-article}.
In the weak coupling regime, when $g_{\gamma} < \kappa$,
the photon losses in the cavity overcome the electron-photon coupling element. 
As a result, photons may leave the cavity faster than being absorbed by the electronic system. 
In the weak coupling regime a high-performance single-photon source achieves very high efficiency and it can be used as a
multi-photon interferometric solid-state device~\cite{PhysRevLett.116.020401}. 

The strong coupling regime opens the way towards a deterministic building of qubit-photon entanglement \cite{PhysRevLett.109.240501},
single photon states~\cite{Houck2007}, and long-range coupling of semiconductor qubits \cite{Majer2007,Sillanpaa2007}. 
In addition, the strong coupling regime is most desirable for optoelectric nanodevices and it is an emerging technology  
in the microelectronics industry in the form of solid state-based quantum processors~\cite{Wu2019}. 

In previous publications, we demonstrated the effects of a photon cavity on the
time-dependent electron transport in a strong electron-photon coupling regimes for both 
charge and thermoelectric transport through, a QD~\cite{Nzar.25.465302,en12061082}, a DQD~\cite{PhysicaE.64.254}, quantum wires~\cite{Nzar_ACS2016,nzar27.015301,Vidar85.075306} and quantum rings~\cite{Abdullah2017}. 
In theses publications, we have shown that 
the photon field in the cavity can be used to control the transport properties of the systems
in the early transient regime were non-Markovian effects may be important.
In the present work, we assume a quantum dot system coupled to a photon cavity with a single 
photon mode in the steady-state regime. We consider a strong coupling between the QD system and 
the photon cavity, ($g_{\gamma} > \kappa$). 
We use a Markovian quantum master equation to investigate the current through the QD system that is generated due to 
photon replica states under the influences of the photon polarization, the cavity-environment coupling strength and the average 
photon number in the environment.

The outline of the paper is as follows. We describe the model system in~\sec{Sec:Model}.
Results are discussed for the model in~\sec{Sec:Results}. Finally, we draw our conclusion in~\sec{Sec:Conclusion}.

\section{Theoretical formalism}\label{Sec:Model}

\begin{figure}[htb]
\centering
 \includegraphics[width=0.45\textwidth]{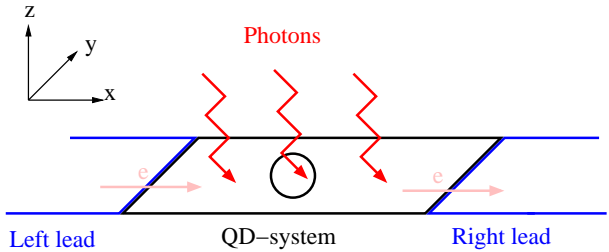}
\caption{Schematic diagram displaying the QD-system (black) connected
 to the leads (blue) and coupled to a photon field (red).}
\label{fig01}
\end{figure}

Our system consists of a quantum dot embedded in a short two-dimensional quantum wire as is 
presented in \fig{fig01}. The QD system (black circle) is connected to two leads (blue) and coupled to a photon 
field (red) in a 3D-cavity with a single photon mode. 

The Hamiltonian describing the QD-system coupled to a photon cavity in the many-body (MB) basis is given as~\cite{2040-8986-17-1-015201,PhysRevE.86.046701,Nzar_2016_JPCM,Nzar.25.465302}  
\begin{equation}
  \hat{H} = \hat{H}_e + \hat{H}_{\gamma} + \hat{H}_{e\text{-}\gamma},
  \label{eq:H}
\end{equation}
with $H_e$ the Hamiltonian of the QD-wire system
\begin{eqnarray}
 \hat{H}_e &=&\int dr\; \hat{\psi}^{\dagger}(\mathbf{r}) \Big[\frac{\bm{\pi}^2}{2m^*} + \frac{1}{2} 
           m^*\Omega_0^2 y^2 + V_{\rm QD} + eV_{\rm g}\Big] \hat{\psi}(\mathbf{r}) \nonumber \\
           & +& H_{\rm Z} + \int dr \int dr' \hat{\psi}^{\dagger}(\mathbf{r})  \hat{\psi}^{\dagger}(\mathbf{r}') V_{c}(\mathbf{r},\mathbf{r}') 
           \hat{\psi}(\mathbf{r}') \hat{\psi}(\mathbf{r}),
           \label{eq:H_e}
\end{eqnarray}
including the electron-electron Coulomb interaction.
Herein, $\bm{\pi} := \mathbf{p}+ (e/c) \mathbf{A}$ with $\mathbf{p}$ being canonical momentum, 
$\mathbf{A} = -By \mathbf{\hat{x}}$ is the magnetic vector potential with $\mathbf{B} = B \mathbf{\hat{z}}$,
and $\hat{\psi}(\mathbf{r}) = \sum_i \psi_i(r) d_i$ and $\hat{\psi}^{\dagger}(\mathbf{r}) = \sum_i \psi_i^*(r) d_i^{\dagger}$ are 
the electron field operators with $d_i$($d^{\dagger}_i$) the annihilation(creation) operators 
for the single-electron state $\ket{i}$ corresponding to $\psi_i$. 
The electron confinement frequency due to the lateral parabolic potential is defined by $\Omega_0$ in 
the short quantum wire and the potential of the QD is described by 
\begin{equation}
  V_{\rm QD} = V_0 e^{(-\gamma_x^2 x^2 - \gamma_y^2 y^2)},
\end{equation}
with $V_0$ its strength, 
and $\gamma_x$($\gamma_y$) are constants that define the diameter of the QD, respectively.
The gate voltage, $V_{\rm g}$, moves the energy states of the QD-wire system with respect to the chemical potential of the 
leads, and it is assumed to be constant in our calculations.
The Zeeman Hamiltonian referring to the interaction between the magnetic moment of an electron and the external magnetic field (B),
is give by $H_{\rm Z}= \pm g^{*}\mu_B B/2$ with $\mu_B$ the Bohr magneton and $g^{*} = -0.44$ the 
effective g-factor for GaAs. 
In addition, the electron-electron interaction is shown in the second line of \eq{eq:H_e} with 
$V_{c}$ being the Coulomb interaction potential~\cite{Nzar.25.465302}. 
The Coulomb interaction in the leads is neglected.

The second term of \eq{eq:H} is the Hamiltonian of the free photon field defined via
\begin{equation}
 H_{\gamma} = \hbar \omega_{\gamma} a^{\dagger} a,
\end{equation}
with $\hbar \omega_{\gamma}$ the energy of the photons in the cavity, and $a$($a^{\dagger}$) the photon annihilation(creation) operators, respectively.

The last part of the \eq{eq:H} stands for the interaction between the electron in the QD system and 
the photons in the cavity
\begin{equation}
 \hat{H}_{e\text{-}\gamma} = \frac{1}{c} \int d{\mathbf{r}} \ \mathbf{j}(\mathbf{r}) \cdot\mathbf{A}_\gamma 
 + \frac{e}{2m^* c^2} \int d{\mathbf{r}} \, \rho(\mathbf{r}) A_\gamma^2,
        \label{Hp-d}
\end{equation}
where the first part of \eq{Hp-d} is the para-magnetic and the second part is the dia-magnetic electron-photon interaction.
The charge density is given by $\rho = -e \psi^{\dagger} \psi$ and the 
charge current density can be introduced by
\begin{equation}
 \mathbf{j} = \frac{e}{2m^*}\left\{\psi^\dagger\left({\bm{\pi}}\psi\right)
                 +\left({\bm{\pi}}^*\psi^\dagger\right)\psi\right\}.
\end{equation}
with $\psi$ the field operator of the QD system.
In addition, the photon vector potential, $\mathbf{A}_{\gamma}$, in the Coulomb gauge is
\begin{equation}
 \hat{\mathbf{A}}_\gamma = A (\hat{a}+\hat{a}^\dagger)\bm{e}. 
\end{equation}
Herein, $A$ is the amplitude of the photon field introduced by the electron-photon coupling constant $g_\gamma=eAa_w\Omega_w/c$. 
The photon polarization can be determined by $\bm{e}$, in which $\bm{e} = \bm{e}_x$ in the $x$-direction 
and $\bm{e}=\bm{e}_y$ in the $y$-direction. We assume the wavelength of the FIR cavity photons to be much larger than
the size of the short quantum wire and the quantum dot.
As a step wise exact numerical diagonalization technique is used to treat the electron-photon and the 
Coulomb interactions interaction in appropriately truncated Fock-spaces, the electron-photon interaction is treated well 
beyond a traditional dipole approximation \cite{Vidar61.305}.

A quantum master equation is utilized to study the transport properties of the system in the steady-state regime 
in which a projection formalism based on the density operator is used \cite{Zwanzing.33.1338,Nakajima20.948}.
The leads and the central system are assumed to be weakly coupled leading to terms of higher than second
order in terms of the coupling to be neglected in the dissipation kernel of the resulting
integro-differential equation. We assume that the QD system and leads are uncorrelated before the coupling 
\begin{equation}
 \hat{\rho}(t < t_0) = \hat{\rho}_\mathrm{L} \hat{\rho}_\mathrm{R} \hat{\rho}_\mathrm{S}(t < t_0),
\end{equation}
where $\hat{\rho}_\mathrm{L}$ and $\hat{\rho}_\mathrm{R}$ are the density operators of the left (L) 
and the right (R) leads, respectively. 

As we are interested in the state of the wire-QD system after the coupling, 
we can obtain the reduced density operator of the QD system from a partial trace over 
the combined QD system and leads
\begin{equation}
 \hat{\rho}_\mathrm{S} = {\rm Tr_\mathrm{ L,R}}[\hat{\rho}(t)].
\end{equation}
The non-Markovian generalized master is
\begin{align}
      \partial_t{\hat{\rho}_\mathrm{S}}(t)  = & -\frac{i}{\hbar}[H_\mathrm{S},\hat{\rho}_\mathrm{S}(t)]
      -\frac{1}{\hbar}\int_0^t dt' K[t,t-t';\hat{\rho}_\mathrm{S}(t')] \nonumber \\
      & -\frac{\kappa}{2\hbar}(\bar{n}_R+1)\left\{2\alpha\hat{\rho}_\mathrm{S}\alpha^\dagger - \alpha^\dagger \alpha\hat{\rho}_\mathrm{S} - \hat{\rho}_\mathrm{S} \alpha^\dagger \alpha\right\}\nonumber\\
       &-\frac{\kappa}{2\hbar}(\bar{n}_R)\left\{2\alpha^\dagger\hat{\rho}_\mathrm{S}\alpha - \alpha\alpha^\dagger\hat{\rho}_\mathrm{S} - \hat{\rho}_\mathrm{S} \alpha\alpha^\dagger\right\},
\label{GME}
\end{align}
where the coupling of the single cavity photon mode is assumed Markovian, and a rotating wave approximation 
has been used only for this coupling.  
The second term of the first line of \eq{GME} describes the electron ``dissipation'' processes caused by both leads,
and the second and the third lines of \eq{GME} represents the photon reservoir where 
$\kappa$ is the photon-cavity coupling constant to the environment (seen as a photon reservoir), 
$\bar{n}_\textrm{R}$ is the mean photon number in the reservoir.
The photon operators in the cavity, $a$ and $a^\dagger$, are replaced by $\alpha$ and $\alpha^\dagger$, that  
lead to the correct steady state by removing all high frequency creation terms from the annihilation operator,
and high frequency annihilation terms from the creation operator
\cite{GUDMUNDSSON20181672,PhysRev.129.2342,PhysRevA.31.3761,PhysRevA.84.043832}.
Subsequently, a Markovian approximation is applied to the master equation (\ref{GME}), 
and a vectorization together with a Kronecker tensor product transforms it from 
the many-body Fock space of photon dressed electron states into a Liouville space of transitions to
facilitate numerical and analytical solutions \cite{JONSSON201781}.  

We assume the chemical potential of the left lead is higher than that of the right lead
producing a bias voltage that generates current through the QD-system coupled to the leads. 
The charge current from the left lead into the QD-system, $I^{c}_{\rm L}$,
and the current from it into the right lead, $I^{c}_{\rm R}$, can 
be introduced as 
\begin{equation}
 I^{c}_\mathrm{L,R} = {\rm Tr}_\mathrm{S} \Big( \Lambda^{\rm L,R}[\hat{\rho}_\mathrm{S};t] Q \Big).
\end{equation}
The charge operator of the QD-system is $Q = -e \sum_i d_i^\dagger d_i$ with $\hat{d}^\dagger (\hat{d})$
the electron creation (annihilation) operator of the central system, respectively.  
$\Lambda^\mathrm{L,R}$ stand for the ``dissipation'' processes caused by both electron 
leads~~\cite{GUDMUNDSSON_2019,JONSSON201781}. The average total number of photons in the
cavity is evaluated as
\begin{equation}
      N_\gamma = {\rm Tr}_\mathrm{S} \Big(\hat{\rho}_\mathrm{S}a^\dagger a \Big) ,
\end{equation}
and each term in the trace operation performed in the basis of interacting electrons 
and photons can be regarded as the photon content of the corresponding dressed electron 
state.

\section{Results}\label{Sec:Results}

The system under investigation is a QD embedded in a short two-dimensional quantum wire 
with hard-wall confinement in the $x$-direction and parabolic confinement in the 
$y$-direction with characteristic confinement energy $\hbar\Omega_0 = 2.0$~meV. The system 
is exposed to a weak external perpendicular magnetic field, $B=0.1$ T, in the $z$-direction leading to 
a cyclotron energy $\hbar \omega_c = 0.172$~meV. Therefore, the effective confinement energy 
is given by $\hbar \Omega_w = \big[ \hbar\Omega_0^2 + \hbar \omega_c^2 \big]^{1/2}$. 
With the weak external magnetic field we avoid the effects of the Lorentz force on 
electron transport in the QD system. The role of this magnetic field is to lift 
the spin degeneracy by a small Zeeman splitting. 

The QD system is connected to two leads with $\mu_L = 1.25$~meV
the chemical potential of the left lead and for the right lead $\mu_R = 1.15$~meV.  
The gate voltage is $V_{\rm g} = 0.651$~meV which is set to place the first photon replica 
of the one-electron ground state in the bias window. 
The temperature of the leads is fixed at $T = 0.5$~K.
The weak external perpendicular magnetic field is also applied to both leads.

The wire-QD system is coupled to a photon field with energy $\hbar \omega_{\gamma} = 1.31$~meV. 
The photon energy is smaller than the energy distance between the one-electron ground state and
the first excitation thereof. Under this condition the QD system is off-resonant with the cavity,
but the anisotropic polarizability of the charge in the system is different for the two linear
photon polarizations along the $x$- and the $y$-direction, as will be seen below. 
We intentionally choose the off-resonant regime, with a photon replica in the bias window to study the current 
yield generated by such states.

\begin{figure}[htb]
  \includegraphics[width=0.23\textwidth,angle=0,bb=60 65 200 300]{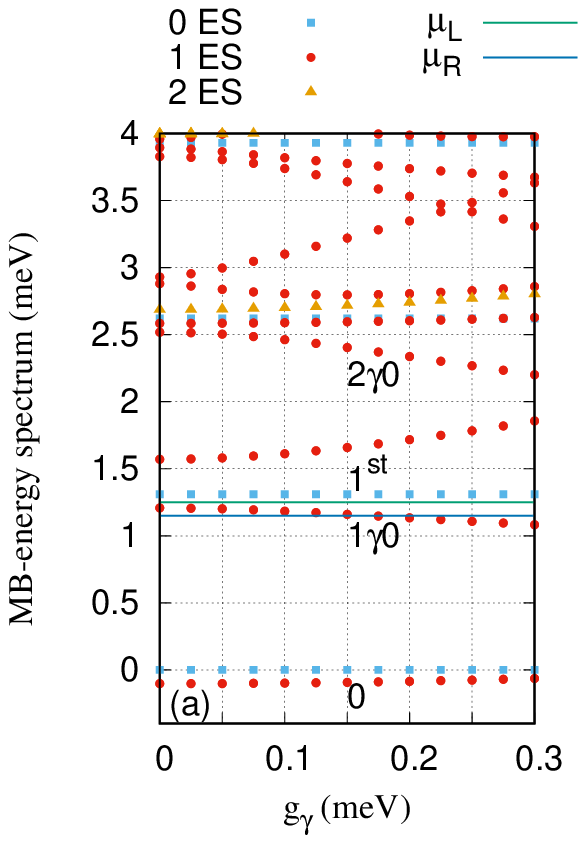}
  \includegraphics[width=0.23\textwidth,angle=0,bb=80 63 222 300]{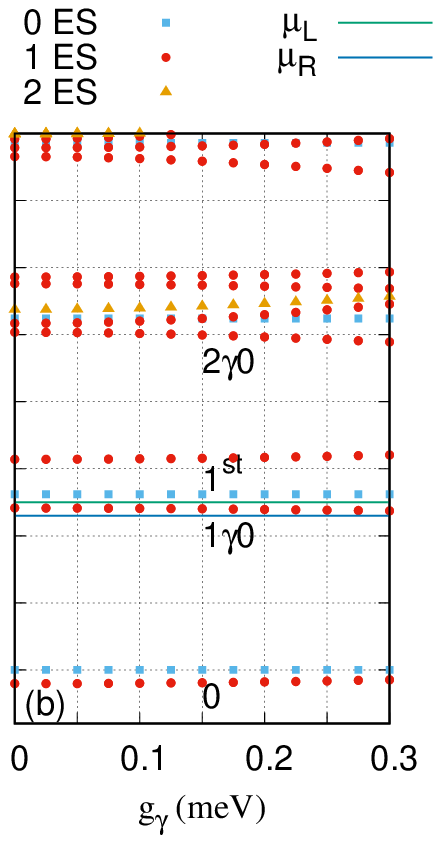}
\caption{Many-Body (MB) energy spectrum ($\rm E_{\rm \mu}$) of the quantum dot system 
        versus the electron-photon coupling strength ($\rm g_{\gamma}$), for $x$- (a) and $y$-polarized (b) photon field,
        where 0ES (blue squares) are zero-electron states, 1ES (red circles) are one-electron states, 
        and 2ES (brown triangles) are two-electron states.
        The chemical potential of the left lead is $\mu_L = 1.25$~meV (green line) and of the the right lead is 
        $\mu_R = 1.15$~meV (blue line).
        $0$ indicates the one-electron ground-state, $1^{\rm st}$ displays the one-electron first-excited state, and 
        $1\gamma$0 and $2\gamma$0 refer to the first and second photon replicas of the one-electron ground-state, respectively.
        The photon energy $\hbar \omega_{\gamma} = 1.31$~meV, $\kappa = 10^{-5}$ meV, and $\bar{n}_\textrm{R} = 1$.
        The magnetic field is $B = 0.1~{\rm T}$, $V_{\rm g} = 0.651$~meV, $T_{\rm L, R} = 0.5$~K and $\hbar \Omega_0 = 2.0~{\rm meV}$.}
\label{fig02}
\end{figure}

Figure \ref{fig02} shows the many-body energy spectrum of the wire-QD system as a function of 
the electron-photon coupling strength, $g_{\gamma}$, for the $x$- (a) and $y$-polarized (b) photon field.
$0$ indicates the one-electron ground-state and $1^{\rm st}$ refers to the first-excited one-electron state.
In addition, $1\gamma$0 and $2\gamma$0 are the first and the second photon replicas of the one-electron ground-state, 
respectively.

At the given gate voltage, $V_{\rm g} = 0.651$~meV, the first photon replica of the ground-state, $1\gamma$0, 
is found to be in the bias window.  Tuning the electron-photon coupling strength, the energy of the states is shifted up or 
down and can form anti-crossings, especially in the $x$-polarized photon field. 
For instance, $1\gamma$0 is shifted down with increasing $g_{\gamma}$ and leaves the bias window 
at high electron-photon coupling strength in the $x$-polarized photon field, while it remains in the bias window for 
the $y$-polarization. In addition, the $1^{\rm st}$ state is approaching $2\gamma$0 starting to form an anti-crossing 
at high electron-photon coupling strength in the $x$-polarization while the same phenomenon can not be seen for the $y$-polarization.
The states are effectively stronger coupled to the $x$-polarized photon field compared to the 
$y$-polarization as the anisotropy of the system makes the charge more polarizable in the $x$-direction.

\begin{figure}[htb]
 \includegraphics[width=0.4\textwidth,angle=0,bb=70 70 410 250]{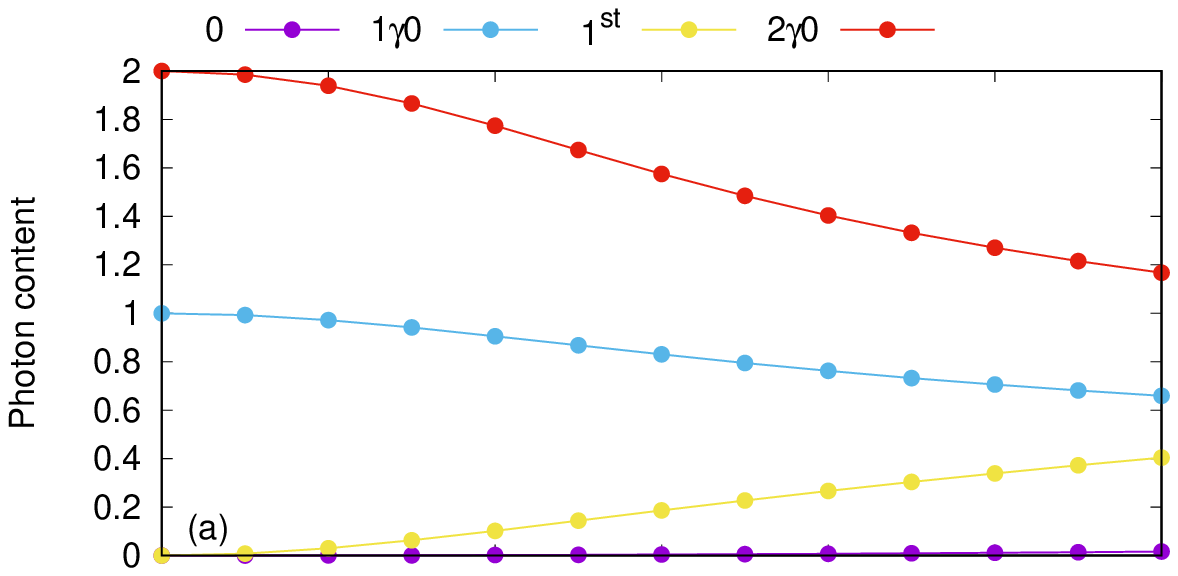}\\
  \includegraphics[width=0.4\textwidth,angle=0,bb=70 55 409 222]{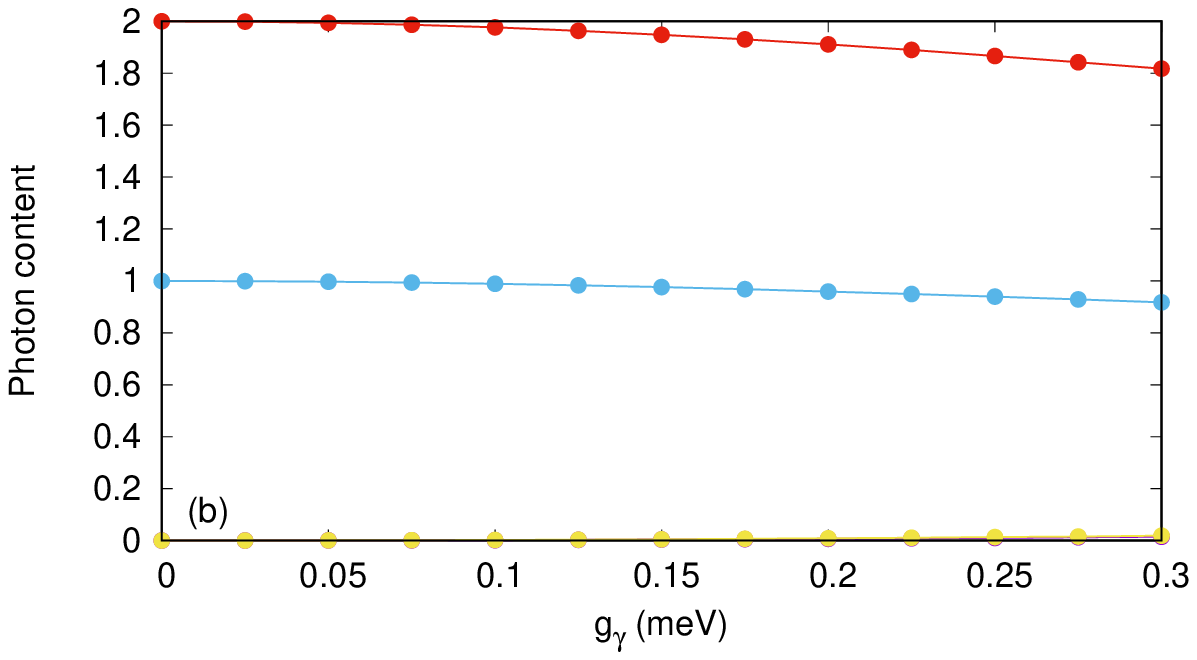}
        \caption{Photon content of the four lowest one-electron states (1ES) as a function of the electron-photon 
        coupling strength ($\rm g_{\gamma}$), 
        for $x$- (a) and $y$-polarized (b) photon field,
        where 0ES (blue squares) are zero-electron states, 1ES (red circles) are one-electron states, 
        and 2ES (yellow triangles) are two-electron states.
        The photon energy $\hbar \omega_{\gamma} = 1.31$~meV, $\kappa = 10^{-5}$ meV, and $\bar{n}_\textrm{R} = 1$.
        The chemical potential of the left lead is $\mu_L = 1.25$~meV and the right lead is $\mu_R = 1.15$~meV.
        The magnetic field is $B = 0.1~{\rm T}$, $V_{\rm g} = 0.651$~meV, $T_{\rm L, R} = 0.5$~K, 
        and $\hbar \Omega_0 = 2.0~{\rm meV}$.}
\label{fig03}
\end{figure}

The mean photon number or photon content of the aforementioned four one-electron states is shown in \fig{fig03} for the 
$x$-polarized (a) and $y$-polarized photon field (b). Increasing the electron-photon coupling strength, 
photon-exchange between the $1^{\rm st}$ and $2\gamma$0 states for the $x$-polarization is observed. The photon content of 
$1^{\rm st}$ is enhanced to $\simeq 0.4$ and the photon content of $2\gamma$0 decreases to $\simeq 1.2$. 
In addition, the photon content of $1\gamma$0 is suppressed to $\simeq 0.65$ in the $x$-polarization.
The characteristics of the photon content here together with the energy spectra shown in \fig{fig02} indicate that 
the states $1^{\rm st}$ and $2\gamma$0 are approaching a Rabi-resonance in the case of an $x$-polarized  
photon field~\cite{Nzar-arXiv_article_2019}.

\begin{figure}[htb]
\includegraphics[width=0.47\textwidth]{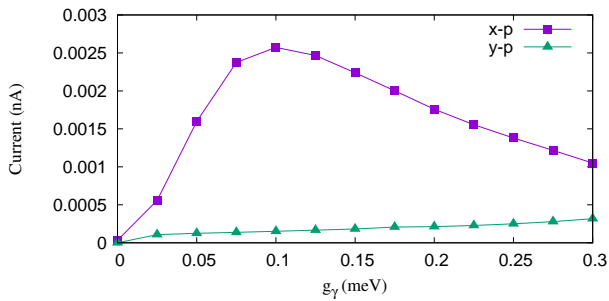}
\caption{Current from the left lead to the QD-system ($I_{\rm L}$)
               as a function of the electron-photon coupling strength, $\rm g_{\gamma}$, for 
               $x$- (purple rectangles) and $y$-polarized (green triangles) of the photon field. 
               The photon energy is $\hbar \omega_{\gamma} = 1.31$~meV, $\kappa = 10^{-5}$ meV, and $\bar{n}_\textrm{R} = 1$. 
               The chemical potential of the left lead is $\mu_L = 1.25$~meV and the right lead is $\mu_R = 1.15$~meV.
               The magnetic field is $B = 0.1~{\rm T}$, $V_{\rm g} = 0.651$~meV, $T_{\rm L, R} = 0.5$~K, 
               and $\hbar \Omega_0 = 2.0~{\rm meV}$.}
\label{fig04}
\end{figure}

We now present the properties of the electron transport displayed by the current through the system.
The current carried by the electrons depends on the width and location of the bias window. 
Here we assume the chemical potential of the left(right) lead to be $1.25$($1.15$~meV), respectively.
The current from the left lead into the QD system is displayed in \fig{fig04} for $x$- (purple rectangles) and 
$y$-polarized (green rectangles) photon field. The current is enhanced with the electron-photon 
coupling strength reaching a maximum at $g_{\gamma} = 0.1$~meV for the $x$-polarized photon field 
while the current is very small and remains almost constant with increasing $g_{\gamma}$ in the $y$-polarization. 
The enhancement of current is related to the charge and the electron-photon dressed states which are more polarizable 
in the $x$-direction. In addition, the transport through the photon replica state such as $1\gamma$0 located in the bias window 
is enhanced with $g_{\gamma}$. As expected the current after $g_{\gamma} = 0.1$~meV, decreases as the 
photon replica state, $1\gamma$0, leaves the bias window. 
We stress that the characteristics of current would not be the same as is shown in \fig{fig04} if there would only be 
slightly photon dressed states of the QD system or additional electron-photon dressed states together with 
the photon replica states in the bias window~\cite{doi:10.1002/andp.201700334}.
We note that the left and the right currents here, i.e.\ the current through the QD,
are equal in magnitude because the system is in the steady-state regime.

\begin{figure}[htb]
 \includegraphics[width=0.45\textwidth,angle=0,bb=70 70 410 250]{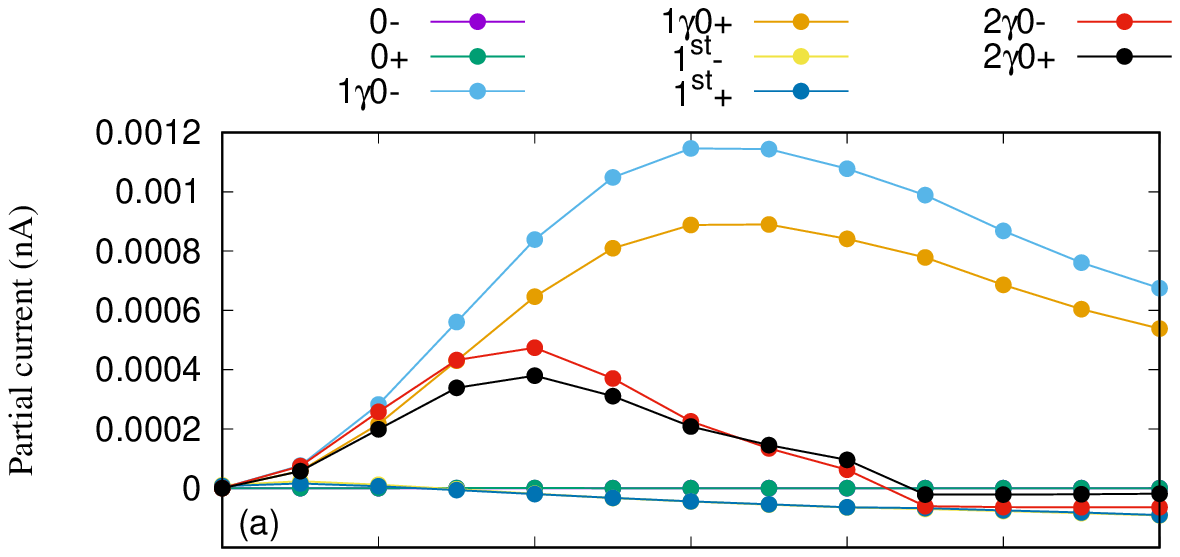}
 \includegraphics[width=0.45\textwidth,angle=0,bb=64 60 410 195]{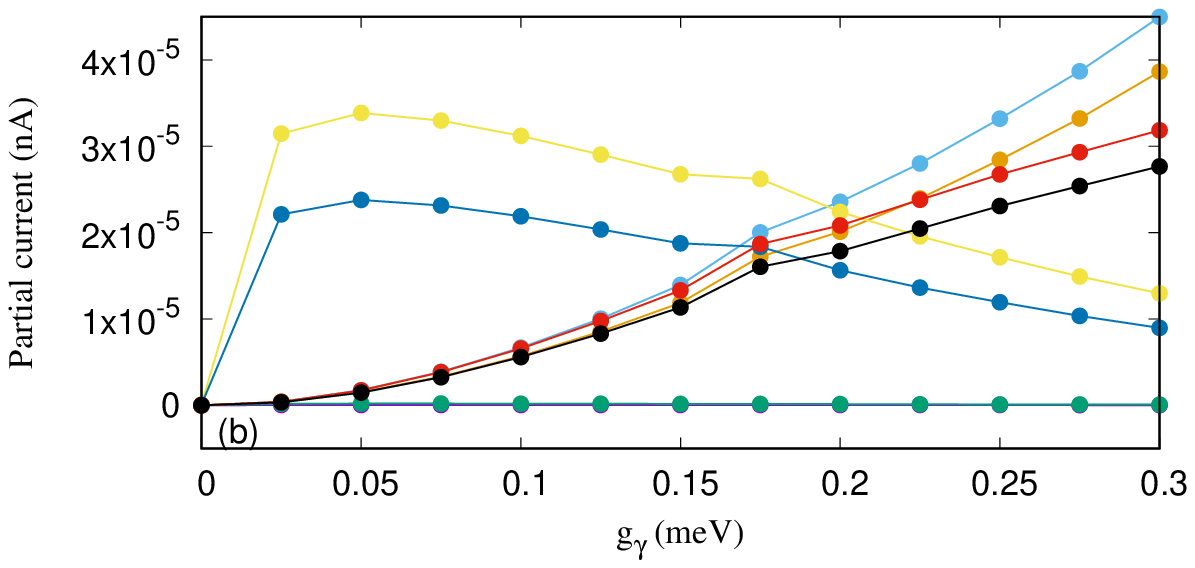}
\caption{Partial current of the four lowest one-electron states versus the electron-photon coupling 
        strength ($\rm g_{\gamma}$) for $x$- (a), $y$-polarized (b) photon field.
        Herein, 0- indicates the one-electron ground-state with spin-down (purple) and 0+ the spin-up (green), 
        $1\gamma$0- refers to the one-photon replica of the $0$ with spin-down (light blue) and $1\gamma$0+ for spin-up (orange),
        $1^{\rm st}$- displays the one-electron first-excited state with spin-down ($-0.5$) (yellow) and $1^{\rm st}$+ 
        the spin-up (dark blue),
        and $2\gamma$0- is the two-photon replica of $0$ with spin-down (red) and $2\gamma$0+ the spin-up (black).
        The photon energy is $\hbar \omega_{\gamma} = 1.31$~meV, and $\kappa = 10^{-5}$ meV, $\bar{n}_\textrm{R} = 1$.
        The chemical potential of the left lead is $\mu_L = 1.25$~meV and the right lead is $\mu_R = 1.15$~meV.
        The magnetic field is $B = 0.1~{\rm T}$, $V_{\rm g} = 0.651$~meV, $T_{\rm L, R} = 0.5$~K, and $\hbar \Omega_0 = 2.0~{\rm meV}$.}
\label{fig05}
\end{figure}

To understand the detailed characteristics of the electron transport we present \fig{fig05} which shows 
the partial currents, i.e.\ the currents going through individual states, for the $x$- (a) and $y$-polarized (b) photon field.
For the given chemical potentials the first photon replica, $1\gamma$0, is located in the bias window.
In $x$-polarized photon field, $1\gamma$0 is the most active state in the transport as is shown in \fig{fig05}(a) in which the 
partial current for the four lowest states of the QD system coupled to the cavity is plotted.
The current through $1\gamma$0- with spin down (light blue) and $1\gamma$0+ for spin up (brown)
is enhanced by increased electron-photon coupling strength up to $g_{\gamma} = 0.15$~meV.
The current is suppressed at higher electron-photon coupling strength in the case of $x$-polarization. 
The reason for the current suppression through $1\gamma$0 after $g_{\gamma} = 0.15$~meV is that 
$1\gamma$0 is moving out of the bias window as the coupling increases. 
In addition to the first photon replica state, the second photon replica state $2\gamma$0 participates in 
the electron motion and a small current through 
$2\gamma$0 is observed for both spin down (red) and spin up (black). Again, the current through $2\gamma$0 for both spin components
is approaching zero but at the same time the first-excited state gains charge leading to generation
of a small current via $1^{\rm st}$ with 
both spin components (yellow and dark blue) at high electron-photon coupling strength, $g_{\gamma}>1.5$~meV.
The discharging of $2\gamma$0 and charging of $1^{\rm st}$ is related to the photon-exchange between 
these two states shown in \fig{fig03}(a) leading to an intraband transition that occurs between them.
The characteristics of total current shown in \fig{fig04} follow the partial currents
of the photon replica states $1\gamma$0 and $2\gamma$0 shown in \fig{fig05}(a). 
It indicates that more than $95\%$ of the total current is generated due to the contribution of 
the photon replica state, $1\gamma$0, to the electron transport. 
Therefore, we can call the total current the photo-generated, or the photocurrent of the QD system. 

We should mention that the current through the aforementioned states for the $y$-polarized photon field
is much smaller, $100$ times smaller, than that for the $x$-polarization as is displayed in \fig{fig05}(b). This confirms that 
the $y$-polarized photon field does not influence much the electron transport and no important effects of the photons 
can be seen in the energy spectrum and the photon content.
This is related to the anisotropy of the QD system in which charge is more polarizable in 
the $x$-direction~\cite{PhysicaE.64.254}. 

\begin{figure}[htb]
  \includegraphics[width=0.45\textwidth]{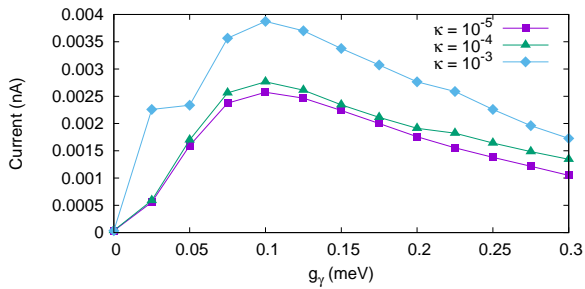}\\
       \caption{Current from the left lead to the QD-system ($I_{\rm L}$)
               as functions of the electron-photon coupling strength $\rm g_{\gamma}$ for 
               $\kappa = 10^{-5}$ meV (purple rectangles), $10^{-4}$ meV (green triangles), and 
               $10^{-4}$ meV (blue diamonds).
               The photon energy is $\hbar \omega_{\gamma} = 1.31$~meV, $\bar{n}_\textrm{R} = 1$,
               and the photon field is linearly polarized in the $x$-direction. 
               The chemical potential of the left lead is $\mu_L = 1.25$~meV and the right lead is $\mu_R = 1.15$~meV.
               The magnetic field is $B = 0.1~{\rm T}$, $V_{\rm g} = 0.651$~meV, $T_{\rm L, R} = 0.5$~K, and $\hbar \Omega_0 = 2.0~{\rm meV}$.}
\label{fig06}
\end{figure}

\begin{figure}[htb]
 \includegraphics[width=0.45\textwidth,angle=0,bb=55 70 410 250]{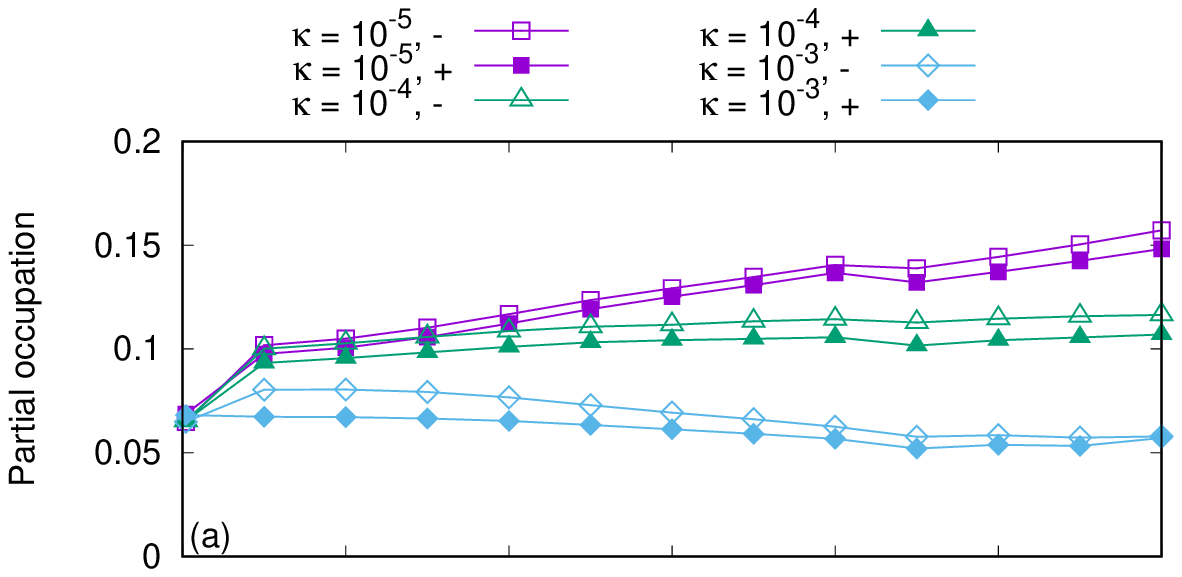}
 \includegraphics[width=0.45\textwidth,angle=0,bb=62 60 409 195]{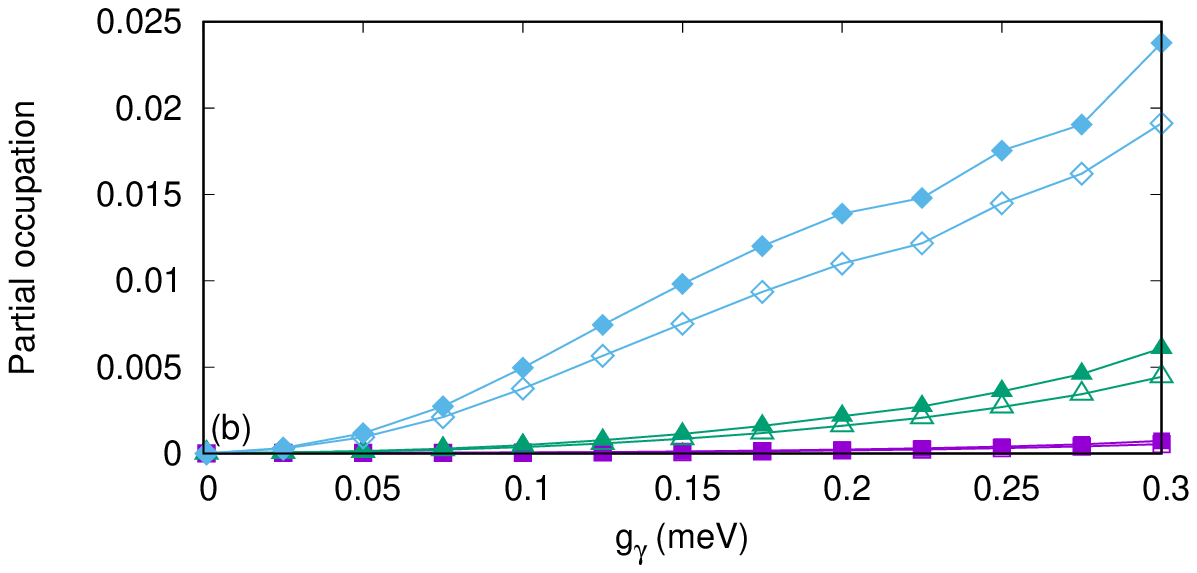}
\caption{Partial occupation of $1\gamma$0 (a) and $1^{\rm st}$ (b) versus the electron-photon coupling 
        strength ($\rm g_{\gamma}$) for $\kappa = 10^{-5}$ meV (purple squares), $10^{-4}$ meV (green triangles),
        and $10^{-3}$ meV (blue diamonds) with spin-down (hollow) and spin-up (solid).
        The photon energy is $\hbar \omega_{\gamma} = 1.31$~meV, $\bar{n}_\textrm{R} = 1$, and
        the photon field is linearly polarized in the $x$-direction. 
        The chemical potential of the left lead is $\mu_L = 1.25$~meV and the right lead is $\mu_R = 1.15$~meV.
        The magnetic field is $B = 0.1~{\rm T}$, $V_{\rm g} = 0.651$~meV, $T_{\rm L, R} = 0.5$~K, and $\hbar \Omega_0 = 2.0~{\rm meV}$.}
\label{fig07}
\end{figure}

\begin{figure}
 \includegraphics[width=0.45\textwidth,angle=0,bb=70 70 410 250]{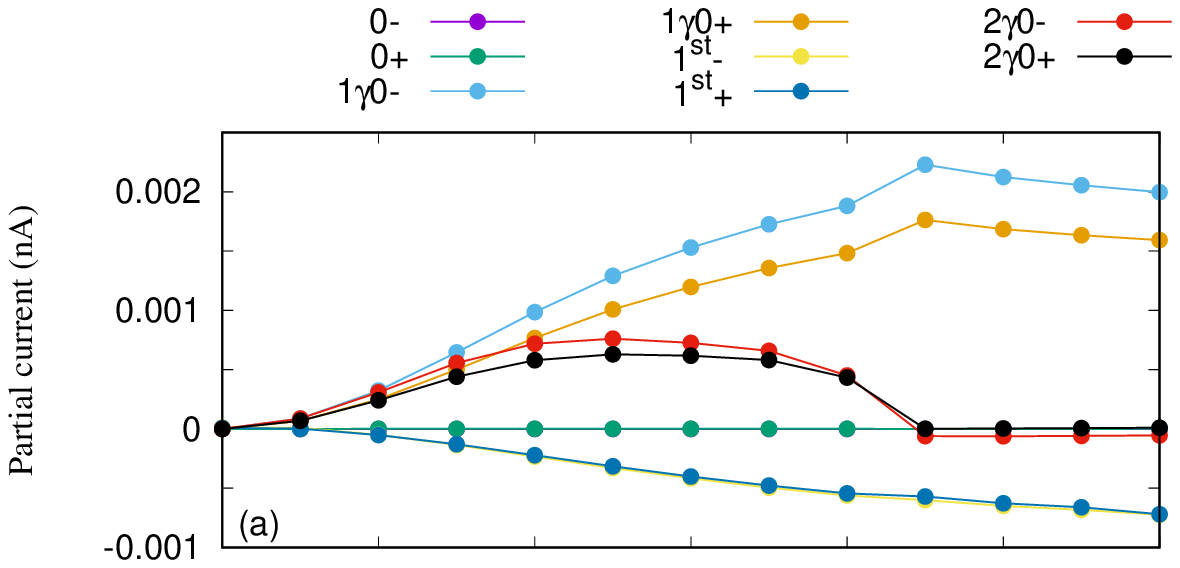}
 \includegraphics[width=0.45\textwidth,angle=0,bb=64 60 410 195]{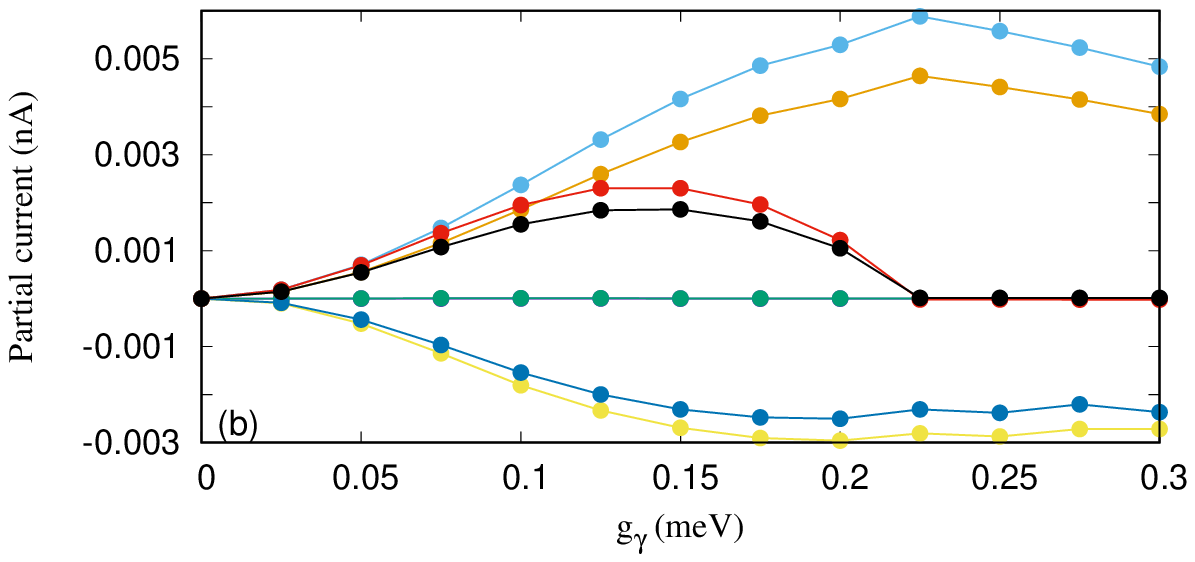}
\caption{Partial current of the four lowest one-electron states (1ES) versus the electron-photon coupling 
        strength ($\rm g_{\gamma}$) for $\kappa = 10^{-4}$ meV (a), and $10^{-3}$ meV (b).
        Herein, 0- indicates the one-electron ground-state with spin-down (purple) and 0+ spin-up (green), 
        $1\gamma$0- refers to the one-photon replica of $0$ with spin-down (light blue) and $1\gamma$0+ spin-up (orange),
        $1^{\rm st}$- displays the one-electron first-excited state with spin-down (yellow) and spin-up $1^{\rm st}$+ (dark blue),
        and $2\gamma$0- is the two-photon replica of $0$ with spin-down (red) and spin-up $2\gamma$0+ (black).
        The photon energy is $\hbar \omega_{\gamma} = 1.31$~meV, $\bar{n}_\textrm{R} = 1$, and
        the photon field is linearly polarized in the $x$-direction. 
        The chemical potential of the left lead is $\mu_L = 1.25$~meV and the right lead is $\mu_R = 1.15$~meV.
        The magnetic field is $B = 0.1~{\rm T}$, $V_{\rm g} = 0.651$~meV, $T_{\rm L, R} = 0.5$~K, and $\hbar \Omega_0 = 2.0~{\rm meV}$.}
\label{fig08}
\end{figure}

We now further study the properties of the current in the case of an $x$-polarized photon field and neglect
the transport properties for the $y$-polarization because the current is too small.
We now tune the cavity-environment coupling $\kappa$, but keep in mind that the electron-photon coupling strength is 
still greater than the cavity-environment coupling, $g_{\gamma} > \kappa$. 
As $n_\mathrm{R}$, the mean value of photons in the reservoir, is not zero the parameter $\kappa$ both influences 
the rate of flow of photons out and into the cavity. 
The current versus the electron-photon coupling strength for different values of $\kappa$ is shown in \fig{fig06}. 
Clearly, the current is enhanced with increasing $\kappa$.
To explain, we refer to the partial occupation and current of the QD system. 
Figure \ref{fig07} demonstrates the partial occupation of the first photon replica of the one-electron ground state,
$1\gamma$0 (a), and the electronic state, $1^{\rm st}$ (b), versus the electron-photon coupling strength. 
We should remember that only $1\gamma$0 is located in the bias window and 
$1^{\rm st}$ is above the bias window (see \fig{fig02}a). 
The occupation of $1\gamma$0 decreases with increasing photon-reservoir coupling rate for all electron-photon coupling strength
(see \fig{fig07}a) while the occupation of $1^{\rm st}$ is enhanced (see \fig{fig07}b). 
This behavior indicates that the participation of photon replica states 
to the transport becomes weak at high photon-reservoir coupling while 
the pure electronic states are populated at the same cavity-reservoir coupling.   
In addition, for low photon-reservoir coupling the most active state is $1\gamma$0 which is due to a 
photon accumulation in the QD system leading to intraband transitions between photon replicas.
The occupation of $1^{\rm st}$ at high photon-reservoir coupling enhances the current through it (until
it moves outside the bias window) as is shown in \fig{fig08} for $\kappa = 10^{-4}$ meV (a) and $\kappa = 10^{-3}$ meV (b).
The current through $1^{\rm st}$ seems to be blocked by the photon cavity for low values of $\kappa$.
As a result the total current through the QD system is enhanced with increasing $\kappa$.

\begin{figure}
 \includegraphics[width=0.45\textwidth,angle=0,bb=70 60 410 250]{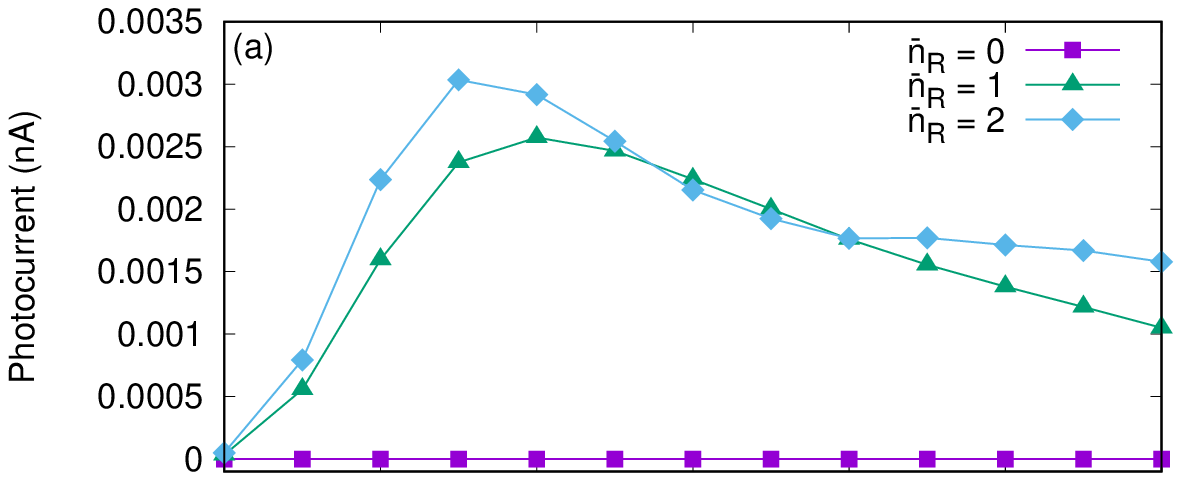}
 \includegraphics[width=0.45\textwidth,angle=0,bb=50 60 410 195]{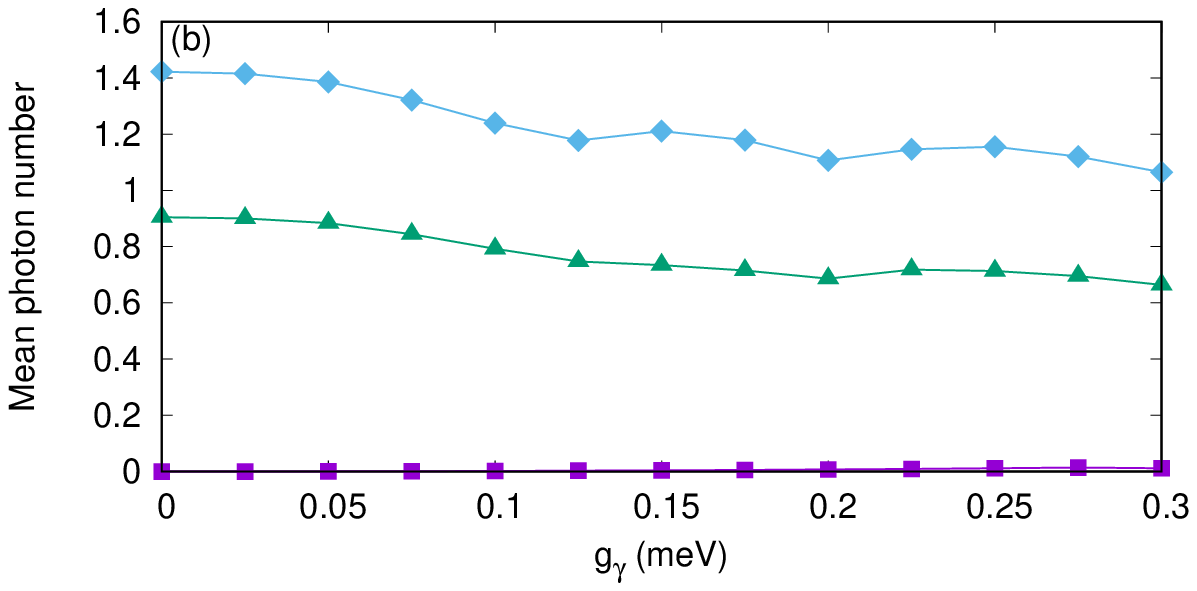}
      \caption{The current from the left lead to the QD-system ($I_{\rm L}$) (a),
      	       and the mean total photon number $N_\gamma$ (b)
               as functions of the electron-photon coupling strength $\rm g_{\gamma}$ for 
               $\bar{n}_\textrm{R} = 0$ (purple rectangles), $1$ (green triangles), and $2$ (blue diamonds).
               The photon energy is $\hbar \omega_{\gamma} = 1.31$~meV, $\kappa = 10^{-5}$ meV,
               and the photon field is linearly polarized in the $x$-direction.
               The chemical potential of the left lead is $\mu_L = 1.25$~meV and the right lead is $\mu_R = 1.15$~meV.
               The magnetic field is $B = 0.1~{\rm T}$, $V_{\rm g} = 0.651$~meV, $T_{\rm L, R} = 0.5$~K, 
               and $\hbar \Omega_0 = 2.0~{\rm meV}$. Note that the lowest value for $g_\gamma$ in the
               figure is $0.001$ meV.}
\label{fig09}
\end{figure}

Another feature of our system is the effect of mean photon number $\bar{n}_R$ in the reservoir on the transport properties.
We assume the cavity-reservoir coupling, $\kappa = 10^{-5}$ meV, and the chemical potential of the leads 
to be fixed as the above calculations. We keep in mind that only  $1\gamma$0 is located 
in the bias window for the given values of the chemical potentials and the gate voltage.
Figure \ref{fig09}(a) shows the current as a function of the electron-photon coupling strength for 
different mean photon numbers in the reservoir. 
The current is almost zero when the mean number of photons is zero, $\bar{n}_\textrm{R} = 0$, 
which is expected because the photon replica states are not active in the transport in the present situation. 
For $\bar{n}_\mathrm{R}=0$ the system enters a Coulomb blocking regime in the steady state.
In this case, the occupation of $1\gamma$0 is almost zero and in turn the current vanishes. 
Clearly, the most occupied state here is the ground state which does not contribute to 
the transferred current through the QD system because it is far below the bias window (see \fig{fig02}). 
If we assume $\bar{n}_\textrm{R} = 2$ (blue diamonds), the contribution of $1\gamma$0 and especially
$2\gamma$0 is slightly enhanced which can be seen from the occupation of these two states (now shown). 
Therefore, the current is slightly enhanced for the case of two photons. This happens as the photon
replicas are not pure simple perturbational states with an integer number of photons, but instead
contain states with 0, 1, and 2 photons at least to some amount.
The total mean photon number $N_\gamma$ displayed in Fig.\ \ref{fig09}(b) is invariably lower than
mean photon number in the reservoir $\bar{n}_R$, as the flow of electrons through the system is maintained
by the ``consumption'' of photons, i.e.\ a photocurrent is maintained in the system. In preparation of 
Fig.\ \ref{fig09} the coupling $g_\gamma$ is never put lower than $0.001$ meV, but the coupling of the cavity
to the environment $\kappa$ is kept constant. For the lowest $g_\gamma$ the system still achieves a steady-state,
but now in a very long time. (We do not set $g_\gamma$ exactly equal to zero as in that unphysical
limit it is technically difficult to account for the approach to the
steady state properly within the numerical accuracy set by the time
scale needed).     

\begin{figure}[htb]
 \includegraphics[width=0.45\textwidth]{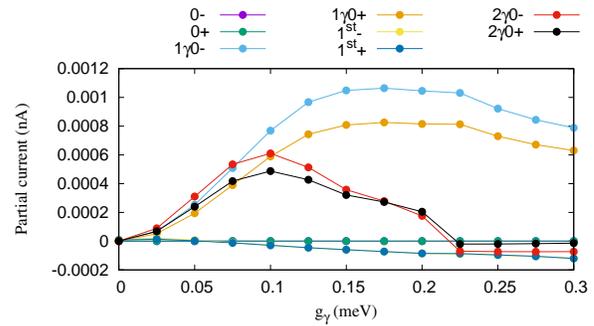}
\caption{Partial current of the four lowest one-electron states (1ES) versus the electron-photon coupling 
        strength ($\rm g_{\gamma}$) for $\bar{n}_\textrm{R} = 2$.
         Herein, 0 indicates the one-electron ground-state with spin-down ($-0.5$) (purple) and spin-up ($0.5$) (green), 
        $1\gamma$0 refers to the one-photon replica of $0$ with spin-down ($-0.5$) (light blue) and spin-up ($0.5$) (orange),
        $1^{\rm st}$ displays the one-electron first-excited state with spin-down ($-0.5$) (yellow) and spin-up ($0.5$) (dark blue),
        and $2\gamma$0 is the two-photon replica of $0$ with spin-down ($-0.5$) (red) and spin-up ($0.5$) (black).
        The photon energy is $\hbar \omega_{\gamma} = 1.31$~meV, $\kappa = 10^{-5}$, and
        the photon field is linearly polarized in the $x$-direction. 
        The chemical potential of the left lead is $\mu_L = 1.25$~meV and the right lead is $\mu_R = 1.15$~meV.
        The magnetic field is $B = 0.1~{\rm T}$, $V_{\rm g} = 0.651$~meV, $T_{\rm L, R} = 0.5$~K, and $\hbar \Omega_0 = 2.0~{\rm meV}$.}
\label{fig10}
\end{figure}

The partial current through the four lowest states is displayed in \fig{fig10} for $\bar{n}_\textrm{R} = 2$.
It can clearly be seen that the current through $1\gamma$0 and $2\gamma$0 is slightly increased in the case of 
two photons leading to a slight enhancement of the total current.

\section{Conclusion}\label{Sec:Conclusion}

We have calculated the transport properties through a quantum dot system connected to leads and 
coupled to a photon cavity with a single photon mode. 
We focus on the transport properties of the photon replica states 
that are formed in the presence of the photon field coupled to the QD system. 
These photon replica states can be confined in the bias voltage by setting the chemical 
potential of the leads. 
In this way, one can see $95\%$ of the current in the system can be obtained
due to the photon replica states. 
We can thus show the influences of photon polarization, mean photon number in the reservoir, and photon-reservoir coupling rate 
on the transport properties in the system. We find that the photon polarization plays an important role and can 
be used to control the photocurrent generated in the system. In addition, the total current is enhanced with 
increasing the photon-reservoir coupling rate because the partial 
current carried by both the almost pure electronic states and the photon replica states is increased.

It is important to have in mind that we have not considered very strong electron photon coupling 
if we compare the photon coupling strength with the energy difference between the one-electron ground
state and the first electronic excitation thereof. We have also not selected the photon energy to be
very close to this energy difference, but as we account for geometrical effects in our anisotropic
system it is clear the that effective electron-photon coupling becomes rather strong, and the only
way to approach this regime sincerely is by using a step wise exact numerical diagonalization scheme for 
all interactions in the central system. Even though, we do consider the variation of the photon fields
to be small on the size scale of the electronic system, we are not using a traditional dipole approximation
and the higher order interaction effects are important in delivering a more appropriate cavity photon dressed
electron states to describe transport of electron through our system. Radiative transitions in our system with
a FIR photon mode take time, and the variation of the cavity-photon reservoir coupling strength, $\kappa$, 
can be used to activate them or reduce their effects.

\begin{acknowledgments}

This work was financially supported by the Research Fund of the University of Iceland,
the Icelandic Research Fund, grant no.\ 163082-051, 
and the Icelandic Infrastructure Fund. 
The computations were performed on resources provided by the Icelandic 
High Performance Computing Center at the University of Iceland.
NRA acknowledges support from University of Sulaimani and 
Komar University of Science and Technology.
CST acknowledges support from Ministry of Science and
Technology of Taiwan under grant No.\ 106-2112-M-239-001-MY3.

\end{acknowledgments}


%

%
\end{document}